\documentclass[prl,nofootinbib,twocolumn,superscriptaddress]{revtex4}

\usepackage{graphicx}

\usepackage{amsbsy}

\def\mysection#1{{\bf #1.} }
\def\mysections#1{{\bf #1.} }

\newcommand{\be}{\begin{equation}}
\newcommand{\ee}{\end{equation}}
\newcommand{\beq}{\begin{eqnarray}}
\newcommand{\eeq}{\end{eqnarray}}
\newenvironment{Eqnarray}%
         {\arraycolsep 0.14em\begin{eqnarray}}{\end{eqnarray}}
\def\beqa{\begin{Eqnarray}}
\def\eeqa{\end{Eqnarray}}

\def\lsim{\mathrel{\rlap{\lower4pt\hbox{\hskip1pt$\sim$}}
    \raise1pt\hbox{$<$}}}         
\def\gsim{\mathrel{\rlap{\lower4pt\hbox{\hskip1pt$\sim$}}
    \raise1pt\hbox{$>$}}}         

\def\eps{\varepsilon_{\rm L}}

\def\eV{{\rm~eV}}

\def\TeV{{\rm~TeV}}

\begin{document}
 
\vspace*{-10mm}

\title{\begin{flushright}{\small
\vspace*{-1.5cm} LBNL-54272}
\end{flushright}\vspace*{.1cm}
{\boldmath Leptogenesis from Split Fermions} 
}

\author{Yukinori Nagatani}\email{yukinori.nagatani@weizmann.ac.il}
\affiliation{Department of Particle Physics,
  Weizmann Institute of Science, Rehovot 76100, Israel}

\author{Gilad Perez}\email{gperez@lbl.gov}
\affiliation{Theoretical Physics Group, Lawrence Berkeley National Laboratory, Berkeley, CA 94720}

\vspace*{1cm}

\begin{abstract}
  We present a new type of leptogenesis mechanism based on a two-scalar split-fermions framework.
  At high temperatures the bulk scalar vacuum expectation values (VEVs)
  vanish and lepton number is strongly violated.
  Below some temperature, $T_c$, the scalars develop extra dimension
  dependent VEVs. This transition is assumed to proceed via a first
  order phase transition. In the broken phase the fermions are
  localized and lepton number violation is negligible. The lepton-bulk scalar
  Yukawa couplings contain sizable CP phases which induce lepton
  production near the interface between the two phases.
  We provide a qualitative estimation of the resultant baryon asymmetry which
  agrees with current observation.
   The neutrino flavor parameters are accounted for by the above model
  with an additional approximate U(1) symmetry.
  \end{abstract}

\maketitle

\mysection{Introduction}
The SM is inconsistent with the following observations:
1.~CP violation
(CPV), where the Standard Model SM sources cannot explain the observed baryon
asymmetry of the universe (BAU).
2.~Neutrino masses, for which
there is strong evidence from various experiments.
In addition the SM raises the  
flavor puzzle: Most of the SM flavor parameters
are small and hierarchical.

In~\cite{GrP} it was shown that the flavor puzzle
is naturally solved within the split fermion framework~\cite{AS}
using a two scalar model.
In this work we focus on the lepton sector and show how the first two puzzles, listed above,
can be address within the same framework.
We demonstrate that it can explain the observed BAU via a new leptogenesis 
mechanism.
Our model realizes the following idea, previously considered
in~\cite{Prev}: At temperatures above the critical one, $T_c$, 
the universe is in the symmetric phase.
In this phase lepton number is strongly violated due to the fact that the
wave functions (WFs) of the lepton zero modes
are flat and have large overlaps between them.
At $T_c$ a bubble of broken phase is formed in which the WFs are localized and
lepton violating interactions vanish.

The essential difference between our work and the previous ones
is that in our case all the required ingredients for baryogenesis are related to
the bulk scalar sector which is an inherent part of the model.
To see this we describe how the
Sakharov conditions~\cite{Sak} are satisfied in the above:
(i) {\it Lepton violation} - Above $T_c$, the bulk scalar
VEVs vanish. The leptons are not localized and 
lepton number is strongly violated~\cite{Prev}.
At $T_c$ a bubble of non-zero VEVs is formed.
In the bubble, unlike in the symmetric phase, lepton number is only violated through sphaleron
processes which translate the lepton excess into a baryon one~\cite{Leptogenesis}.
(ii) {\it C,CP violation} - The fermion-bulk scalar Yukawa couplings 
are generically of order one and 
they contain an order one CPV phase. Thus, unlike in the SM,
CPV interactions with the bubble wall are unsuppressed by a small
Jarlskog determinant~\cite{Jarlskog,HuSa}.
(iii) {\it Deviation from equilibrium} -
The transition between the broken and the unbroken phase is assumed to occur via a 
first order phase transition (PT).

Our model also accounts for the neutrino flavor parameters using
an additional approximate U(1) lepton  
symmetry as described below.

The model requires a flat extra dimension, $|x_5|\leq\pi R$, 
compactified on an orbifold, $S_1/Z_2$,  where the bulk scalars
are odd with respect to the $Z_2$.
The work in~\cite{GrP} is extended to account for the neutrino 
flavor parameters which are induced by a variant of the minimal
seesaw model~\cite{Ms,RS}.
Our model consists of  $L^i$, SU(2) lepton doublets,
${\ell}^i$, charged leptons, where $i$ is a flavor index, $N_{1,2}$
SM singlet neutrinos (in principle an additional singlet neutrino can
  be added provided that its couplings to the SM fields are
  sufficiently suppressed) and, $\Phi_{1,2}$, the SM singlets bulk scalars.
The relevant part of the Lagrangian is given by:
\beq
{\cal L}={\cal L}_{\rm Y}+{\cal L}_{\rm L}+V_{\Phi,T}\,,
\eeq
where ${\cal L}_{\rm Y}$ contains the interactions between the
leptons and the bulk scalars. 
${\cal L}_{\rm L}$
contains the lepton violating interactions and the Yukawa
couplings to the SM Higgs. $V_{\Phi,T}$ is the bulk scalar temperature
dependent potential which drives the first order PT.

The actual mechanism of creating the asymmetry is similar to the
SM electroweak baryogenesis case. At $T_c\sim {R^{-1}}$ a bubble is formed.
Almost immediately it fills the whole
compact extra dimension and expands in the 4D direction.
Incoming leptons from the unbroken phase
hit the bubble wall. Reflection asymmetries and lepton violating interactions 
induce an excess of leptons which is then overtaken by the bubble.

We first
focus on ${\cal L}_{\rm Y}$ and 
derive a qualitative estimate for the excess of
leptons near the bubble wall. This is first done under the assumption that the rate
for lepton violation, $\Gamma_{\rm L}$, is infinite (zero) in the
unbroken (broken) phase.
Then we consider ${\cal L}_{\rm L}$ which accounts for the observed lepton
flavor parameters.
We calculate $\Gamma_{\rm L}^{-1}$ and find that it is much longer than
other dynamical time scales relevant to our model. This induces a further suppression in
the resultant asymmetry. We then derive a qualitative estimation of
$n_{\rm B}/s$ which agrees with the observed value.
We also briefly comment
on the requirement from $V_{\Phi,T}$ as to yield a first order PT.

\mysection{Interaction with the bubble}
The interaction between fermions of each representation of the SM, $\Psi=L,\ell,N$, and the bubble wall
are given by
\beq
{\cal L}_{\rm Y}={1\over \sqrt {M}}\left(f^{ij}_1 \Phi_1+f^{ij}_2\Phi_2\right) \bar \Psi^i\Psi^j
\label{Y5D}
\,,
\eeq
where $f^{ij}_{1,2}$ are representation dependent hermitian matrices and 
and $M$ is the fundamental scale of the above 5D effective theory.
By a unitary rotation $f^{ij}_1$ can be brought to a real diagonal form
which preserves a $U(1)^3$ global
symmetry.
A-priori, $f^{ij}_2$ contains three phases. Two
phases  can be eliminated using part of the above
$U(1)^3$ symmetry [the whole expression in (\ref{Y5D}) is invariant
under a residual U(1) flavor symmetry]
and thus a single physical CPV phase, $\varphi$, is present.
Generically, all the elements of $f^{ij}_{1,2}$ are of order one.
Consequently, the asymmetry between a process and its CP conjugate
one induced by $f^{ij}_{1,2}$ is expected to be sizable.
This is in clear contrast with the SM baryogenesis case, where
CP asymmetries induced by the Higgs bubble are
tiny due to the smallness of the Jarlskog determinant~\cite{HuSa}.

For our mechanism to work
$T_c$ should be of the order of ${R^{-1}}$ as follows:
At low temperatures, $T\ll{R^{-1}}$, the theory is
essentially 4D, in particular the Yukawa interactions with the bulk scalars
are absent. Thus, $T_c\ll {R^{-1}}$
is unacceptable.
On the other hand, at  $T\gg{R^{-1}}$, thermal processes will
induce lepton violating interaction even in the broken phase which is unacceptable.
Thus we must have
$T_c\sim{R^{-1}}\,.$~\footnote{Constraints on the reheating
  temperature in this case are not very stringent and may be
  evaded~\cite{TRH}.}
This implies that, on-shell, high KK modes are 
irrelevant since their statistical weights are
exponentially suppressed.
Thus, for our purpose, it is enough to consider only
the first few KK modes. In practice, 
we shall consider only the zero and the first KK modes.

Since we are only interested in having a qualitative estimation of the BAU, we
shell not provide
a complete finite temperature analysis of the dynamics near the bubble wall.
We shall apply the naive thin wall 
approximation. This approximation~\cite{HuNe} 
can overestimate the resultant asymmetry. One should therefore view our final result as an
upper bound.
Note, however, that when using
results from the 4D case~\cite{CKNRev} for the values of the mean
free path, $l^{-1}_{\rm Th}\sim g_s^2 T_c$, the velocity of the wall,
$v_{\rm w}\sim0.1$ and the wall width, $\delta_{\rm w}\sim T^{-1}_c$,
then the required conditions for the approximation are met.

As explained above, in this part we consider the case
where there is no suppression due to inefficient lepton
violating interaction. The resultant excess of lepton number, $\bar
n_{\rm L}$, is given by~\cite{HuSa,CKNRev}:
\begin{eqnarray}
\bar n_{\rm L}&=&-\sum_{n,m=0,1^\pm}\int {dp_3\over 2 \pi}n^{\rm u}[E(n,p_3)]\Delta(E)
\,, 
\label{nL}
\end{eqnarray}
where $n^{\rm u}[E(n,p_3)]$ stands for Fermi-Dirac distribution
function for an $n$ KK state with 4D momentum $p_3$ in 
the unbroken phase in the bubble wall rest frame.
$\Delta(E)={\rm Tr}\left(R_{nm}^{\dagger}R_{nm}- \bar
  R_{nm}^{\dagger}\bar R_{nm}\right)\,,$ where
$R_{nm}$ stands for the reflection coefficients which are matrices in
flavor space.
For example, $R_{nm}^{ij}$ is related to an incoming $n=0,1^\pm$ KK lepton state
($n^\pm$ are the two helicity states) of a flavor $i$ which is reflected into an  $m=0,1^\pm$ KK state 
(conserving angular momentum) of flavor $j$. $\bar R_{nm}$
corresponds to the $CP$-conjugate processes.
 Using CPT the expression for $\bar n_{\rm L}$ (\ref{nL}) is further 
simplified~\cite{CKN,HuSa}:~\footnote{We do not apply the quasi-particle
treatment~\cite{Wel,HuSa}.
Unlike the SM case, the asymmetry is created even in the absence of
these thermal effects.}
\begin{eqnarray}
  \bar n_{\rm L}\simeq&-&\int {dp_3\over 2 \pi}
    \left\{ n^{\rm u}[E(0,p_3)]- n^{\rm u}[E(1^+,p_3)]\right\}\Delta(E)\,,
\nonumber\\
\,\Delta(E)&\simeq&{\rm Tr}\left(
     R_{01^+}^{\dagger}R_{01^+}-  R_{1^+0}^{\dagger}R_{1^+0}\right)\,.
\label{nL1}
\end{eqnarray}
To first order ${v}_{\rm w}$ we get~\cite{HuSa}
\begin{eqnarray}
  \bar n_{\rm L}&=&-{{v}_{\rm w}\over T}\int_{1\over R}^{2\over R} {dE\over 2 \pi}
  \left(E-\sqrt{E^2-{R^{-2}}}\right) \times\nonumber
  \\
  &&n_0(E)\left[1-n_0(E)\right]
   \Delta(E)\,,
\label{nL2}
\end{eqnarray}
where $n_0(E)$ is a Fermi-Dirac distribution function in an unboosted
frame and  the integral is taken over the region which yields
the dominant contribution.
From (\ref{nL2}) we learn that our problem is reduced into
finding $\Delta(E)\,.$
In principle, the calculation of $\Delta(E)$ within
the thin wall approximation is straightforward.
One should solve the Dirac eq. for the fermions in the broken and
unbroken phase regions. Then by matching the WFs at the
bubble wall one can extract the reflection matrices.
In practice the required analysis, even numerically,
is very hard.
Thus, we shall apply the following approximation:
We compute the reflection coefficients
in a single generation framework. We expect
that the value of the reflection coefficients will not change
significantly while promoting the model into a three generation one.
The essential difference is that in the latter case
unsuppressed CPV sources are present.
Consequently,  $R_{01^+}^{\dagger} R_{01^+}$
will pick up an order one phase relative to its CP conjugate. 
Qualitatively, we therefore expect the following: 
\beq
\Delta(E)\sim \left|R_{01^+}\right|^2 \sin\varphi\sim \left|R_{01^+}\right|^2
\label{main}\,,
\eeq
where $R_{01^+}$ is calculated in a single flavor model.
Even in this case the effective bulk scalar background
(\ref{Y5D}) has
a complicated structure~\cite{GrP} and the Dirac eq. in the bubble cannot be
solved analytically. As a rough approximation for the
effective VEV we thus take it to be of a step function shape,
$ \langle\Phi\rangle={v\over \sqrt R}\,\left[\theta(x_5)-\theta(-x_5) \right]
\,.
$

In this case the fermion WFs 
can be found analytically in the whole region.
They are characterized by a  spatial 4D momentum, $k_3$ and a KK index $n$.
As angular momentum perpendicular to the wall is preserved,
it is enough to consider only a single (negative) helicity state~\cite{CKN}.
Here we give only the $x_5$ dependent part of the WFs.
In the unbroken phase the WF is of the form:
\beq
\Psi^{\rm u}_{-}(k_3,n)=
\pmatrix{
  N^+_n\cos\left(n{ x_5\over R}\right) \cr
  N^-_n\sin\left(n{ x_5\over R}\right)}\,,
\label{DSolN}
\eeq
where $E^2={n^2\over R^2}+k_3^2$ and $N^\pm_n$ are normalization constants.
Due to the orbifold transformation assignment 
the upper [lower] component in (\ref{DSolN}) which corresponds to a 
left [right] handed field is described by an
even [odd] function.
Including also virtual (off shell) modes for the reflected WF,
the generic solution is a linear combination of the above functions.
In the broken phase, the zero mode WF is given by:
\beq
\Psi^{\rm b}_{-}(k_3,0)=
\pmatrix{
  N'^+_0 e^{fv x_5} \sqrt{2E}\cr
  0}\,,\label{DSolP0}
\eeq
and the other KK states are described by: 
\beq
\Psi^{\rm b}_{-}(k_3,n)=
\pmatrix{
 {N'_n}^+\left[ \cos\left(n{ x_5\over R}\right)+
{fv\over k_3}\sin\left|n{ x_5\over R}\right|\right] \cr
  {N'_n}^-\sin\left(n{ x_5\over R}\right)}\,,
\label{DSolPn}
\eeq
where $E^2=f^2 v^2+{n^2\over R^2}+k_3^2$. 
In split fermion models~\cite{GrP,GP} one typically finds
$f v R\gsim 10$.
In the broken phase, therefore, only the zero mode can be produced on shell.

$R_{01^+}$ is found by matching
the wave function of an incoming zero mode with momentum $k_3$
onto generic WFs of reflected and transmitted fermions.
To match the above WFs, inclusion of the off-shell modes is required
and the actual matching was done numerically.
To test our calculation we verified that we get a zero reflection coefficient
for $ER<1$. 
In addition we checked that unitarity and CPT are satisfied 
by considering also the case of an incoming first KK mode. For this,
we also computed
$R_{1^-1^+}$ and the corresponding transmission coefficients (a more
detailed analysis, including the generalization to three generations, 
will be presented elsewhere).

\begin{figure}[!t]
\begin{center}
\includegraphics[height=4.5cm, width=8cm]{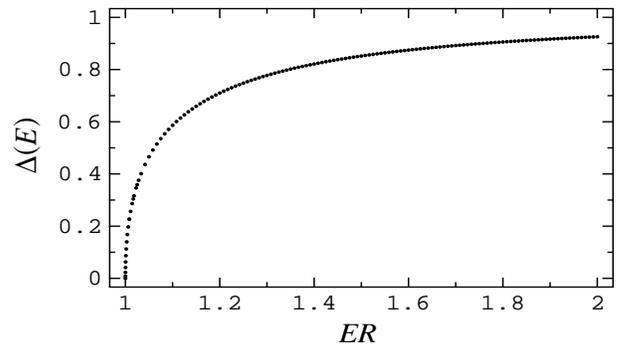}
\caption{$\Delta(E)$ in a single flavor model  for $1<ER<2$.
}\label{figRE}
\end{center}
\end{figure}

The results of our analysis are shown in fig. \ref{figRE}
where we plot $\left|R_{01^+}\right|^2$ for energies in the relevant range, $1<ER<2$. 
The figure shows that a sizable reflection is found over most of the energy range.
Using the result for $\Delta(E)$ we 
numerically performed the integral (\ref{nL2}) and find
\beq
\bar n_{\rm L}\sim 10^{-2}\times v_{\rm w}\label{nL3}\,. 
\eeq
\vspace*{-0.0cm}
\mysection{Lepton flavor sector}
To naturally realize the minimal seesaw scenario~\cite{RS}
we imposed an additional U(1) lepton symmetry.
It is assumed to be broken by a small parameter, $\eps$, which
controls the amount of lepton violation at all temperatures.
Consider the following charge assignment for the leptons under the
U(1)$_{\rm L}$, $
Q(L)=Q(\ell)=3\,,\ \ Q(N)=1\,.$
The relevant part of the 5D Lagrangian is given by 
\beq {\cal
L}_{\rm L}={{\eps^6}\over  {M}}\, LLHH+{\eps^2\over\sqrt  {M}} N^\dagger L H+{\eps^2}  {M}
NN
\label{LH}\,, \eeq
where we suppressed the flavor indices and dimensionless coefficients.

The model yields a hierarchical pattern for
the neutrino masses~(see e.g. \cite{Dat}):
\beq
m_{\rm atm}\sim 5\times10^{-2}\eV^2\,, \ \ \
m_{\rm sol}\sim 8\times10^{-3}\eV^2\,, 
\eeq
and the third neutrino mass is smaller.
We require that the masses induced by the bare term ($L^2 H^2$)
are small, say below $m_{\rm sol}/5$. This implies an upper bound on $\eps$
\beq
\eps\sim 0.055\left( { R^{-1}\over5000\TeV}
\right)^{{1\over6}} 
\label{eps}
\,,
\eeq
with $MR\sim100$ and
$R^{-1}\gsim 5000\TeV$~\cite{FCNC}. 

Given (\ref{LH}) and (\ref{eps}) we can
compute the strength of lepton flavor
violation at $T_c$ in the unbroken phase.
An example for lepton violating process is $NL\to L^\dagger \ell$
which is mediated by a
Higgs, $t$-channel, exchange between $L$ and $N$.
The amplitude for this a process is 
$A_{\rm L}\propto \eps^2 \left({\eps^2 M/ T_c}\right)^2\,,$
where we used the fact that at $T_c$ 
the heavy neutrinos are dynamical~(\ref{LH},\ref{eps}), $M_{N}\sim\eps^2 M\lsim
T_c$. The typical inverse time scale for lepton coherent production near the wall
is~\cite{HuSa,CKN} $\tau^{-1}_{\rm w}\approx v_{\rm w}/l_{\rm w}\sim
v_{\rm w}/T_c$. Thus, the lepton violating rate
is much smaller than $\tau_{\rm w}^{-1}$,
\beq \Gamma_{L} \tau_{\rm w}\approx \left|A_{\rm L}\right|^2
T_c\tau_{\rm w} \sim  {\eps^{12}\over  v_{\rm w}} \left({M\over
    T_c}\right)^4\sim 10^{-8}\,.\label{GamL}
\eeq
In this particular model, therefore,
lepton violating interactions are inefficient in converting the
excess of lepton into anti-lepton, near the wall, before it is
overtaken by the bubble.
Thus, the result in (\ref{nL3}) 
overestimated the resultant excess of leptons.

Let us briefly describe how the neutrino flavor parameters are accounted for in our model.
We denote the required suppression of the 4D neutrino Dirac masses due to the
small overlaps, as $\epsilon_{\rm sol,atm}$~\cite{RS}.
$\epsilon_{\rm sol}$ ($\epsilon_{atm}$) characterize the
overlaps between $N_{1}$ ($N_{2}$) and $L_{1,2}$ ($L_{2,3}$) while
the other overlaps are smaller.
Using the experimental data we
find:~$\epsilon_{\rm atm}\,,\epsilon_{\rm sol}\sim (0.02,0.007)\,.$
This pattern is yielded by the
following bulk scalar-lepton Yukawas, using the notation
of~\cite{GrP} (assuming for simplicity that the Yukawas are
 flavor diagonal), 
\beq
\label{Msbar}
&&f_1^{L_{1,2,3}}=12,-28,15\,,\ \  
  X^{L_{1,2,3}}=-1,0.7,1\,,\nonumber\\
&&f_1^{N_{1,2}}=\pm50\,,\ \  
  X^{N_{1,2}}=0.55,0.55\,,\ \ a_1=3\,.
\eeq
\mysection{Final results}
For our crude estimation of $n_{\rm L}$ we use the
relation $n_{\rm L}\sim \Delta n_{\rm L}^2
\Gamma_{\rm L}\tau_{\rm w} \bar n_{\rm L}$~\cite{CKNRev} 
where  $\Delta n_{\rm L}=2$
is the number of leptons produced by a single lepton violating
process.
Using $(\ref{nL3},\ref{GamL})$ we find
\beq
{n_{\rm L}\over s}&\sim& \Delta n_{\rm L}^2
\Gamma_{\rm L}\tau_{\rm w} {\bar n_{\rm L}\over s}\sim 10^{-10}
\,,\label{final}
\eeq
where we used the relation $s\simeq 2\pi^2 g^*/45\sim 45$.
This is in agreement with the observed value, $n_{\rm B}/s={\cal O}\left(10^{-10}\right)$.

Let us also briefly list the requirements from $V_{\Phi,T}$ 
as to have a first order PT (in order to check whether the requirement below are realistic
a finite temperature analysis is needed which is left for a future work).
We assume that the dominant part of $V_{\Phi,T}$ is given by:
\beq
V_{\Phi,T}={a^2\over 4M^4}\,\phi^6 +{b\over M}\,\phi^4
+{c} M^2\phi^2
\,,\label{V}
\eeq
where $a,b,c$ are real and $c$ is strongly temperature dependent.
A first order PT  implies that at $T_c$ the true and the false vaccua are
separated by a barrier.
This requires that $b<0$ and that $c$ is a monotonic increasing
function of T  
with $c(T_c)\sim{b^2\over a^2}\,$.

\mysection{Conclusions}
In this work
we presented a new class of baryogenesis models within the split fermions framework. 
We focused on the lepton sector and 
demonstrated how our mechanism works.
Note that the final value we
got for the BAU, ${\cal O}\left(10^{-10}\right)$, 
is much smaller than our naive expectation~(\ref{nL3}), ${\cal O}\left(10^{-2}\right)$.
The extra suppression is model dependent and comes from our particular
realization of the minimal see-saw model using the approximate U(1) lepton symmetry, which is not
inherent to the above framework.
Furthermore, we believe that the above mechanism is quite general and can be also applied
to other split fermion models related to the quark sector~\cite{GrP,AS}.

\mysections{Acknowledgments}
G.P. thanks Zacharia Chacko, Walter Goldberger,
Markus Luty and Carlos Wagner for useful discussions and
the Aspen Center of Physics in which
part of this work was done.
The authors thank Roni Harnik, Yuval Grossman and Yossi Nir for useful discussions
and comments on the manuscript.
G.P. is supported by the Director, Office of Science, Office of High
Energy and Nuclear Physics of the US Department of Energy under
contract DE-AC0376SF00098; Y.N. is supported by the Koshland Postdoctoral Fellowship of
the Weizmann Institute of Science.

\vspace*{-5mm}

\end{document}